\documentclass[aps,prl,reprint,groupedaddress]{revtex4-1}

\usepackage{graphicx}  
\usepackage[caption=false]{subfig}
\usepackage{amsmath,amsfonts,amsthm,amssymb}	
\usepackage{mathptmx}
\newcommand{\parallele}
{\protect\scalebox{0.8}{//}}

\usepackage{xfrac} 
\DeclareCollectionInstance{plainmath}{xfrac}{mathdefault}{math}
{
scale-factor = 0.9,
denominator-bot-sep = -1 pt,
numerator-top-sep = -1.5 pt
}
\UseCollection{xfrac}{plainmath}

\newcommand{\gtX}{g^{\scriptscriptstyle(2\text{X})}}
\newcommand{\TR}{T_\text{\;\!R}}
\newcommand{\TL}{T_\text{\ \!L}}
\newcommand{\TA}{T_\text{A}}
\newcommand{\TB}{T_\text{B}}
\newcommand{\TAB}{T_\text{A\,(B)}}
\newcommand{\RA}{R_\text{A}}
\newcommand{\RB}{R_\text{B}}
\newcommand{\RAB}{R_\text{A\,(B)}}
\newcommand{\BSA}{\text{BS}_\text{A}}
\newcommand{\BSB}{\text{BS}_\text{B}}
\newcommand{\BSAB}{\text{BS}_\text{A\,(B)}}
\newcommand{\unit}[1]
{~\text{#1}}
\newcommand{\units}[1]
{\text{#1}}

\newcommand{\sfigHspace}{\hspace{1cm}}

\newcommand{\F}
{\mathrm{F}}

\begin{document}

\title{Measuring the photon coalescence time-window in the continuous-wave regime for resonantly driven semiconductor quantum dots}

\author{Raphael Proux}
\author{Maria Maragkou}
\author{Emmanuel Baudin}
\author{Christophe Voisin}
\author{Philippe Roussignol}
\author{Carole Diederichs}
\email{carole.diederichs@lpa.ens.fr}

\affiliation{Laboratoire Pierre Aigrain, \'Ecole Normale Sup\'erieure - PSL Research University, CNRS, Universit\'e Pierre et Marie Curie - Sorbonne Universit\'es, Universit\'e Paris Diderot - Sorbonne Paris Cit\'e, 24 rue Lhomond, 75231 Paris Cedex 05, France}


\begin{abstract}

We revisit Mandel's notion that the degree of coherence equals the degree of indistinguishability by performing Hong-Ou-Mandel- (HOM-)type interferometry with single photons elastically scattered by a cw resonantly driven excitonic transition of an InAs/GaAs epitaxial quantum dot. We present a comprehensive study of the temporal profile of the photon coalescence phenomenon which shows that photon indistinguishability can be tuned by the excitation laser source, in the same way as their coherence time. A new figure of merit, the coalescence time window, is introduced to quantify the delay below which two photons are indistinguishable. This criterion sheds new light on the interpretation of HOM experiments under cw excitation, particularly when photon coherence times are longer than the temporal resolution of the detectors. The photon indistinguishability is extended over unprecedented time scales beyond the detectors' response time, thus opening new perspectives to conducting quantum optics with single photons and conventional detectors.

\end{abstract}

\maketitle

Indistinguishable photons are one of the keys for the implementation of scalable quantum information systems \cite{QComp,QCrypt}. Indistinguishability is investigated using the coalescence phenomenon: two photons with similar spectral, spatial and polarization properties will bunch when arriving simultaneously on two opposite sides of a beam splitter. One of the pioneers of photon coalescence, Mandel, stated in 1991 that the degree of coherence equals the degree of indistinguishability \cite{Mandel, MandelRev} by investigating theoretically the interference of two light sources, thus underlining the fundamental link of the wave-particle duality of light.

In a two-photon interference Hong-Ou-Mandel (HOM) experiment \cite{Mandel87}, the photons from two sources are combined at the two inputs of a beam splitter and the coalescence will be detected through a drop of the coincidence rate at the outputs---the HOM dip. Under pulsed excitation, perfect temporal matching between the arrival times of the photons at the beam splitter will allow for the observation of the HOM dip. When working with a two-level system, its coherence time $T_{2}$ and its lifetime $T_{1}$ are tightly linked to the photon indistinguishability. For example, perfect coalescence giving rise to a zero value in the HOM dip is observed only if the radiative limit $T_2=2T_1$ is reached. The figure of merit under pulsed excitation is thus given by the ratio $T_2/2T_1$ which constitutes a fundamental limit to the coalescence efficiency \cite{Abram}. Under continuous wave (cw) excitation, with two ideal ultrafast detectors, the coincidence rate always vanishes at zero time delay, even for deviations in the properties of the photons \cite{Legero}. In the case of real detectors, the indistinguishability is thus properly resolved only if the temporal resolution of the detectors $\TR$ is shorter than the coherence time of the photons \cite{Bev}. If $\TR\sim T_1, T_2$, the HOM dip is strongly affected and will disappear completely in the limit of very slow detectors. With a cw source, the value at zero delay of the coincidence rate is thus very sensitive to $\TR$ and does not accurately characterize the intrinsic properties of the source with regard to photon indistinguishability. A new figure of merit has to be considered.

\begin{figure*}
\centering
\includegraphics[width=0.8\textwidth]{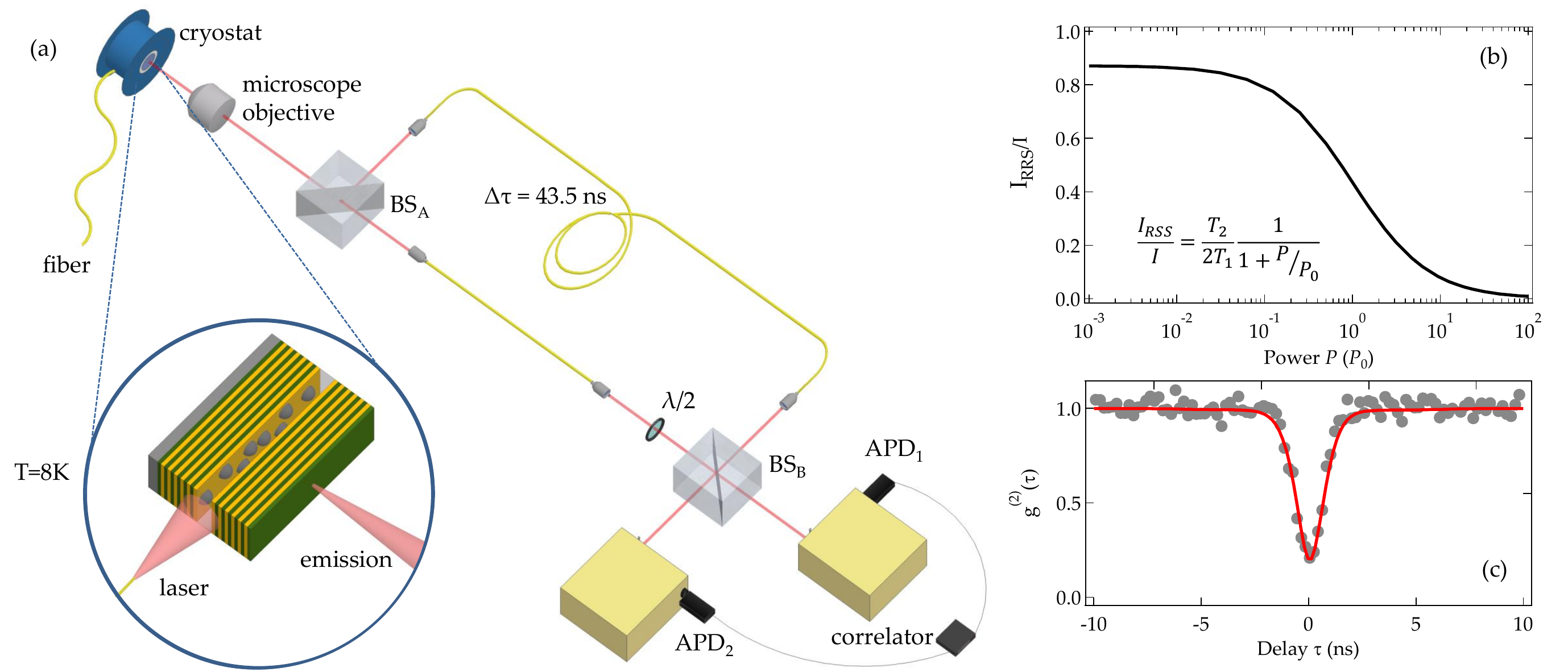}\\
\caption{(a) Unbalanced Mach-Zehnder interferometer for two-photon interference measurements. The QD emission is split by a first beam splitter ($\BSA$) in two paths of different lengths and recombined at a second one ($\BSB$). Single mode polarization maintaining fibers ensure high spatial overlap at $\BSB$ and a half-wave plate $\lambda/2$ controls the mutual polarization between the two interferometer arms. Photodetection is monitored by two avalanche photodiodes (APD) placed at the $\BSB$ outputs, combined with spectrometers for spectral filtering and a correlator for the intensity correlation function measurements. (Inset) Orthogonal excitation-detection geometry where a QD is excited via a fiber positioned at the edge of the sample while its emission is collected by a microscope objective in an orthogonal configuration. (b) Power dependence of the RRS intensity ratio with respect to the total QD emission intensity. The power $P$ is given in units of the saturation power of the excitonic transition $P_0$, $T_1=0.30\unit{ns}$ and $T_2=0.50\unit{ns}$. (c) Intensity correlation function for $P=0.1\,P_0$, fitted by the intensity correlation function of a resonantly excited two-level system \cite{SupInf}, convoluted by the HBT instrument response function (IRF).}
\label{fig1}
\end{figure*}

Single semiconductor quantum dots (QD) \cite{YamamotoNat}, along with other systems under extensive study including atoms \cite{Legero, HOMatomsGrangier}, molecules \cite{HOMmolKiraz, HOMmolLettow, HOMmolLettowRev}, trapped ions \cite{HOMionsInnsbruck, HOMionsMaunz}, and colored centers in diamond \cite{HOMcolCent}, are promising candidates for sources of single indistinguishable photons. In the case of semiconductor QDs, photon indistinguishability is either limited by the QD dynamics under pulsed excitation \cite{Abram}, or by the detectors' temporal resolution under cw excitation \cite{ToshibaPRL,ToshibaNatPhot}. Recent experimental studies focused on the regime of resonant Rayleigh scattering (RRS) under low power cw excitation, where the incoming photons are elastically scattered. This is a well-known phenomenon described by the two-level system resonance fluorescence theory \cite{TBScully}, which has been observed with QDs \cite{Haison,ataturePRL108,Konthasinghe2012,AtatureNatComm,Solomon2013}. As predicted by theory and shown by homodyne and heterodyne detection experiments \cite{AtatureNatComm,Konthasinghe2012,Solomon2013}, the scattered photons inherit the coherence time of the excitation laser $\TL$, which can be much longer than $T_2$ and  $\TR$, while still exhibiting sub-Poissonian statistics \cite{Haison,ataturePRL108}. The resulting QD emission spectrum can then be much narrower than the natural linewidth imposed by the radiative limit, even if the percentage of elastically scattered photons remains limited to $T_2/2T_1$ [Fig.~\ref{fig1}(b)]. Considering that under such conditions the inherited coherence time surpasses $\TR$, along with Mandel's notion that coherence equals indistinguishability \cite{Mandel}, the RRS regime constitutes the ideal ground for the generation of highly indistinguishable photons.

In this Letter, we report on the coalescence of photons emitted by a cw resonantly driven single QD, under this RRS regime. The coherence and indistinguishability properties are defined by the laser coherence time which thus becomes a free controllable parameter of the device. HOM experiments show that photon indistinguishability can be extended to unprecedented time scales and driven externally by the excitation source, without being limited anymore by the QD dynamics or the detection system time response. Furthermore, we revisit the way of estimating the indistinguishability of a cw single photon source by introducing a new figure of merit, the {\it coalescence time window} (CTW), to quantify the delay below which photon coalescence occurs. This criterion sheds new light on the interpretation of HOM experiments under cw excitation of any two-level system, which was traditionally restricted to the evaluation of the two-photon interference visibility at zero delay. Our results also highlight the duality between coherence and indistinguishability, first suggested by Mandel \cite{Mandel}, and provide a novel way of conducting quantum optics experiments by overcoming the limited response function of the detectors.

Our system consists of a single InAs/GaAs self-assembled QD embedded in a planar $\lambda_0$-GaAs microcavity \cite{Haison}. The fundamental excitonic transition is excited resonantly in an orthogonal excitation-detection geometry \cite{Haison,NguyenPRL}, using a cw tunable external cavity diode laser with variable coherence time \cite{SupInf}. The QD emission is sent to a Mach-Zehnder (MZ) interferometer [Fig.~\ref{fig1}(a)]. In order to prevent fictitious anticoincidences from one-photon interference when performing two-photon interference \cite{SupInf}, the path difference must be larger than the photon coherence time, and thus larger than $\TL$ in the RRS regime. This is ensured using optical fibers to reach an interferometer delay $\Delta\tau=43.5\unit{ns}$. A half-wave plate in one of the arms is used to change the polarization. This allows us to make the polarization parallel or orthogonal between the two arms, thus establishing or destroying the interference at the output of the interferometer, respectively. By simply blocking one arm, the setup becomes a Hanbury Brown-Twiss (HBT) setup for measuring the intensity correlation function $g^{(2)}(\tau)$, where $\tau$ is the delay between the detections of the photons. Figure~\ref{fig1}(c) presents the latter at an excitation power $P$ well below the saturation power of the two-level system $P_0$, fitted by the theoretical $g^{(2)}$ \cite{SupInf}. An antibunching dip is observed with $g^{(2)}(0)=0.2$ ($>0$ due to the time resolution of the detectors), demonstrating that the QD is a single photon source in the RRS regime. The QD exciton lifetime $T_1=0.30\unit{ns}$ and the coherence time $T_2=0.50\unit{ns}$ are independently measured under resonant excitation \cite{SupInf}.

\begin{figure}[h!]
\centering
\includegraphics[width=0.4\textwidth]{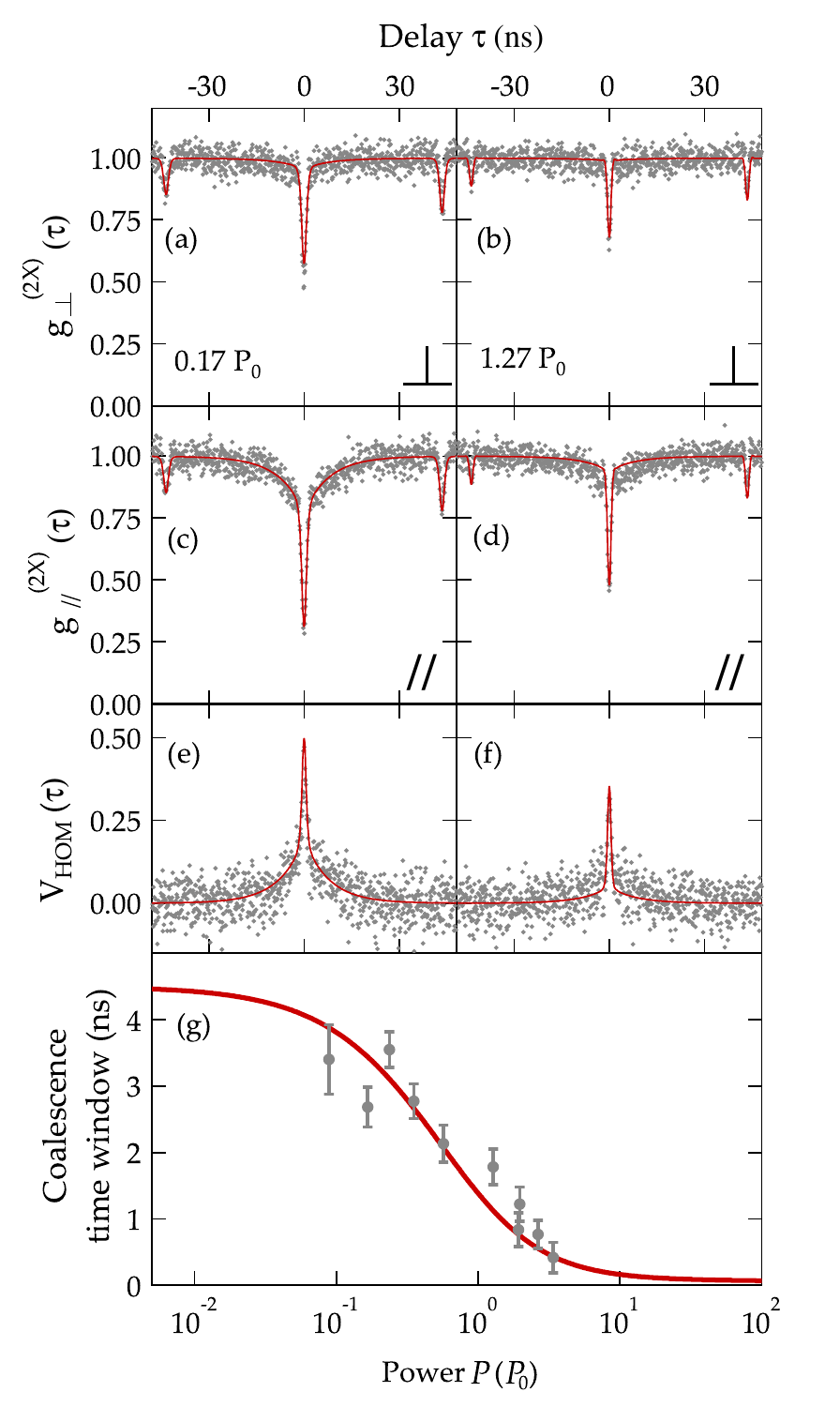}
\caption{(a,b) Two-photon interference measurements for orthogonal polarization configuration ($\perp$) below saturation ($P=0.17\,P_0$) and above saturation ($P=1.27\,P_0$), respectively. (c,d) Same as (a,b) for parallel polarization configuration ($\parallele$). (e,f) Two-photon interference visibility at $P=0.17\,P_0$ and $P=1.27\,P_0$, respectively. The experimental data (dots) are fitted (line) by Eq.~\eqref{g2Xperp} for (a,b), Eq.~\eqref{g2Xpara} for (c,d) and $V_{\text{HOM}}(\tau)$ for (e,f). All the fits are convoluted by the IRF. (g) Coalescence time window deduced from the experimental (dots) and theoretical (line) visibilities as a function of the power $P$. The laser coherence time is $\TL=16\unit{ns}$.}
\label{fig2}
\end{figure}

Figures~\ref{fig2}(a-d) present intensity correlation measurements realized with the HOM setup with orthogonal and parallel polarization configurations (first and second rows, respectively), at low and high excitation powers (left and right columns, respectively), and $\TL=16\unit{ns}$. The experimental data are fitted by the theoretical intensity correlation functions $\gtX_\perp$ and $\gtX_{\parallele}$ \cite{ToshibaPRL}, convoluted by the IRF of the HOM setup of a FWHM of $\TR=1\unit{ns}$ \cite{SupInf}. For the orthogonal polarization configuration,
\begin{multline}\label{g2Xperp}
\gtX_\perp(\tau)=\frac{1}{N}\left[ 4(\TA^2+\RA^2)\RB\TB g^{(2)}(\tau)\right. \\
\left.+ 4\RA\TA\left(\TB^2g^{(2)}(\tau-\Delta\tau)+\RB^2g^{(2)}(\tau+\Delta\tau)\right)\right]
\end{multline}
where $N= 4\RA\TA(\RB^2+\TB^2)+4\RB\TB(\RA^2+\TA^2)$ is a normalizing factor, $\RAB$ and $\TAB$ are the reflection and transmission intensity coefficients of the beam splitters $\BSAB$ respectively, and $\Delta\tau$ is the delay between the two paths of the MZ interferometer. For the parallel polarization configuration,
\begin{multline}\label{g2Xpara}
\gtX_{\parallele}(\tau) = \frac{1}{N}\left[4(\TA^2+\RA^2)\RB\TB g^{(2)}(\tau)\right. \\
+ 4\RA\TA\left(\TB^2g^{(2)}(\tau-\Delta\tau)+\RB^2g^{(2)}(\tau+\Delta\tau)\right) \\ \left.-4\RA\TA(\TB^2+\RB^2)\,V_0\,\big|g^{(1)}(\tau)\big|^2\right]
\end{multline}
where an additional term accounting for two-photon interference appears, with a parameter $V_0$ including all experimental imperfections that destroy the overlap in space or polarization of the two beams at $\BSB$. Here, both equations \eqref{g2Xperp} and \eqref{g2Xpara} depend on the second-order correlation function $g^{(2)}$, while the first-order correlation function $g^{(1)}$, which is linked to the coherence of the two-level system, appears only in Eq. \eqref{g2Xpara} as part of the two-photon interference term. This already highlights that coherence and coalescence are dual notions, as is further investigated below. Note also that because both $g^{(1)}$ and $g^{(2)}$ depend on $T_1$ and $T_2$ ($T_L$ appearing only in $g^{(1)}$) \cite{SupInf}, the dynamics of $\gtX_\perp$ and $\gtX_{\parallele}$ are significantly different from the nonresonant case \cite{ToshibaPRL}. We stress that the same set of parameters has been used for every fit: $\RAB=0.45$ and $\TAB=0.55$; $T_1=0.30\unit{ns}$ and $T_2=0.50\unit{ns}$; $\Delta\tau=43.5\unit{ns}$. Regarding $V_0$, its value is $V_0=0.8$ ($0.15$) for parallel (orthogonal) polarization. The discrepancy between these extracted values and the theoretical ones [$V_0=1(0)$ for parallel (orthogonal) polarization] comes from the spatial mode mismatch and the nonperfect degree of mutual polarizations between the interfering photons. These are mostly due to the use of combined free space and fibered optics, nonpolarizing beam splitters which introduce a small polarization ellipticity, and the gratings of the spectrometers which have a polarization response that partly reestablishes interferences in the orthogonal configuration.

In orthogonal polarization configuration [Fig.~\ref{fig2}(a,b)], no interference is expected and the measured $\gtX_{\bot}$ function is related to the statistical properties of the single photon source when light is sent through the MZ interferometer. Compared to an HBT experiment, additional antibunching dips shifted by the interferometer path difference are observed at $\tau=\pm\Delta\tau$, and the three measured dip values are resolution limited and determined by the QD intrinsic times $T_1$ and $T_2$. At high power [Fig.~\ref{fig2}(b)], the QD undergoes Rabi oscillations \cite{SupInf}, inducing a narrowing of the antibunching dips \cite{Flagg09} and thus a strong reduction of their visibility for a given $\TR$.

In parallel polarization configuration [Fig.~\ref{fig2}(c,d)], in addition to the contribution of the photon statistics, a component due to photon coalescence is observed. At low power [Fig.~\ref{fig2}(c)], two dynamics can be distinguished: a fast one at $\tau\ll \TR$ characterized by the intrinsic QD time constants $T_1$ and $T_2$, and a much slower one characterized by the excitation laser coherence time $\TL$. More specifically, the fast dynamics reflects the photon statistics and the coalescence of the inelastically scattered photons, whereas the slower one is directly linked to the coalescence of the elastically scattered photons. Consequently, these measurements exhibit very clearly two time scales linked to the elastic and inelastic components. At high power [Fig.~\ref{fig2}(d)], the ratio of the elastically scattered photons drops [see Fig.~\ref{fig1}(b)], inducing the long time component to get notably attenuated. These results constitute a clear demonstration of the direct link between the additional coalescence component and the $g^{(1)}$ function, and thus the coherence of the emitted photons. Furthermore, studying the photon indistinguishability in the particular RRS regime provides a straightforward evidence of Mandel's notion regarding the duality between coherence and indistinguishability \cite{Mandel}.

Figures~\ref{fig2}(e,f) present the two-photon interference visibilities $V_{\text{HOM}}(\tau)=\big[\gtX_{\bot}(\tau)-\gtX_{\parallele}(\tau)\big]\big/\gtX_{\bot}(\tau)$ at low and high power, respectively. The usual way to assess the indistinguishability of the photons is to use $V_{\text{HOM}}(0)$. However, this value is heavily altered by the time resolution of the detectors $\TR$. In order to take into account the long coherence time of the elastically scattered photons along with the visibility at $\tau=0$, a more appropriate figure of merit has to be considered: the time integration of the visibility curve, or what we call the CTW. This value should be used under cw excitation in order to investigate the temporal behavior of the coalescence efficiency. This CTW is equal to a weighted average time which takes into account all the temporal components of the coalescence dynamics and therefore corresponds to a relaxation time beyond which no two-photon interference will be observed, while being independent of $\TR$. Figure~\ref{fig2}(g) shows the power dependence of the CTW. At low power (below saturation), it can be as large as $4\unit{ns}$ due to the long coherence of the photons inherited from the excitation laser in the RRS regime. When the power increases, the CTW is drastically reduced and goes below $1\unit{ns}$ above the saturation power. In this regime, the QD emission mostly originates from inelastic scattering, governed by the intrinsic time constants $T_1$ and $T_2$ which are of the order of $\TR$. As a comparison, the CTW calculated for a nonresonantly excited QD with the same time constants $T_1$ and $T_2$ equals $0.15\unit{ns}$, similarly to the one measured at high power (i.e.\ $0.4\unit{ns}$ at $P=5\,P_0$). Here, we conclude that photon coalescence can occur for time delays up to 4 times the detectors' temporal resolution when the QD is operated in the RRS regime (with $\TL=16\unit{ns}$), thanks to the slow dynamics achievable in this particular regime.

\begin{figure}[h!]
\centering
\includegraphics[width=0.4\textwidth]{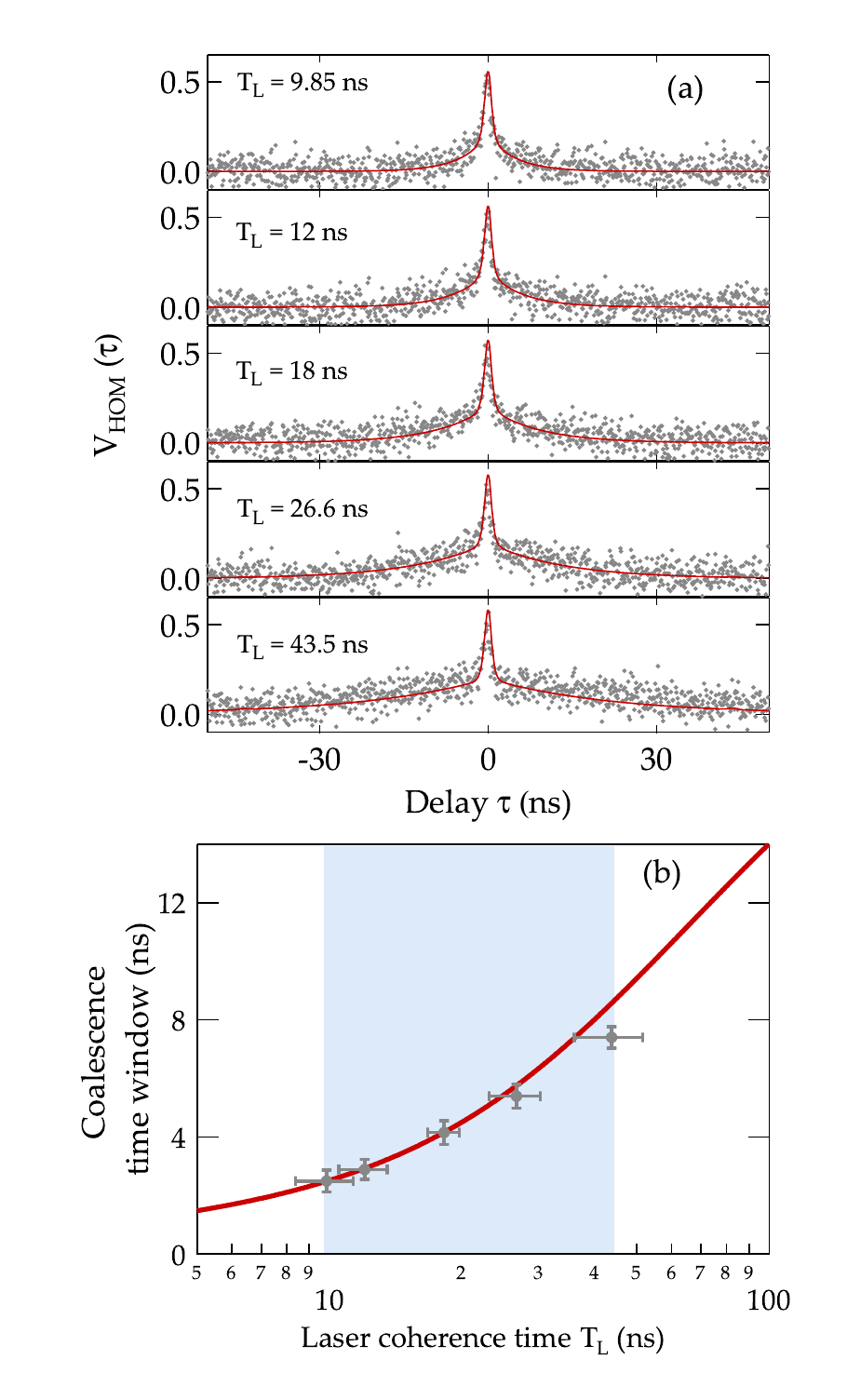}
\caption{(a) Two-photon interference visibility for various laser coherence times $\TL$, at $P=0.3\,P_0$. The experimental data (dots) are shown with the fits (line) given by $V_{\text{HOM}}(\tau)$, convoluted by the IRF. (b) Coalescence time window as a function of $\TL$. The shaded area depicts the reachable range of $\TL$.}
\label{fig3}
\end{figure}

Figure~\ref{fig3}(a) presents two-photon interference visibilities when the QD is in the RRS regime, at low power ($P=0.3\,P_0$), for various laser coherence times. The experimental conditions set the reachable range of $\TL$, between $9.8\unit{ns}$ due to the limitations of our diode laser \cite{SupInf}, and $43.5\unit{ns}$ imposed by the interferometer delay $\Delta\tau$. At a given excitation power, the ratio of the elastically scattered photons remains constant ($I_\text{RRS}/I\approx70\%$), and the increase of $\TL$ is directly reflected on the slow component of the visibility. The corresponding experimental CTWs are presented as a function of $\TL$ in Fig.~\ref{fig3}(b). The theoretical CTW can be calculated within the present model only if $\TL>T_1,T_2$ (domain of validity of the rotating wave approximation). The observed increase of the CTW is directly related to the increase of $\TL$, resulting in a value up to $8\unit{ns}$ with the current setup. Therefore, in the case of RRS, photon indistinguishability as measured by the CTW is limited neither by the intrinsic QD time constants nor $\TR$, and the higher the laser coherence time, the higher the CTW. In addition, the ratio $T_2/2T_1$ only gives the proportion of elastically scattered photons. Consequently, this result not only demonstrates that coherence and indistinguishability are entwined, but also that the RRS regime allows for an unprecedented level of control of photon indistinguishability.

In this Letter, we demonstrate the generation of highly indistinguishable single photons from a cw resonantly driven QD operated in the RRS regime. As the excitation laser drives the photon coherence time beyond the intrinsic properties of the two-level system, the temporal dynamics of the photon coalescence phenomenon can be experimentally investigated. We define an appropriate figure of merit in order to quantify the time window in which two-photon coalescence is observed. This CTW fully characterizes the photon temporal indistinguishability of a cw single photon source, particularly in the RRS regime where photon coherence times are much longer than the temporal resolution of the detectors. We further show that the CTW can be tuned by the excitation laser in the RRS regime and can be as large as $8\unit{ns}$ in the present setup, compared to $0.15\unit{ns}$ for a nonresonantly excited QD, or $0.4\unit{ns}$ for a resonantly driven QD at high power (above saturation). We point out that this new experimental degree of freedom can promote conducting quantum optics with conventional detectors. More specifically, it could be used in a Franson interferometer for the generation of time-bin entangled states out of two single photons under cw excitation \cite{BeveratosENtanglingPhotonsBS}. In such a scheme, as the CTW exceeds the temporal resolution of regular detectors, a precise timing of the photons could be ensured without using narrow spectral filters or superconducting detectors, which would then allow implementing time entanglement of photon pairs \cite{Zukowski_PRL93,BevNature}.


\begin{acknowledgments}

The authors gratefully acknowledge P. Petroff for providing the sample, as well as A. Beveratos and E. del Valle for useful and stimulating discussions. This work was financially supported by the French "Agence Nationale de la Recherce" (ANR-11-BS10-010) and "Direction G\'en\'erale de l'Armement" (DGA).

\end{acknowledgments}

\clearpage
\onecolumngrid

\setcounter{figure}{0}
\renewcommand\thesubfigure{(\alph{subfigure})}

\renewcommand{\thefigure}{S\,\arabic{figure}}

\begin{center}
\large\textbf{{\sffamily Supplemental material:}\\ Measuring the photon coalescence time-window in the continuous-wave regime for resonantly driven semiconductor quantum dots}
\end{center}

\section{S1. Laser diode and Laser Coherence Controller}

The excitation laser is an external-cavity monomode continuous wave diode laser (Toptica DL pro 940). The losses in the cavity and the longitudinal mode selection are ensured by a grating in Littrow configuration which can be rotated to tune roughly the emission wavelength between $910\unit{nm}$ and $985\unit{nm}$. The cavity length can be finely tuned thanks to a piezoelectric actuator. The overall accuracy of the spectral relative position is less than $5\unit{MHz}$, two orders of magnitude below the linewidth of the quantum dot (QD).

The electrical current passing through the diode laser can be modulated by a noise generator (Toptica Laser Coherence Controller -- LCC). The LCC output is a white noise of electrical current with a bandwidth in the $10-250\unit{MHz}$ range, and variable modulation power. The diode laser output is thus modulated in phase and amplitude, resulting in a broadened lorentzian spectral line.

The coherence time of the laser emission is here defined as the decay time $T_L$ of the exponential in the first-order correlation function $g^{(1)}$:
\begin{equation}
g^{(1)} (\tau) = e^{-\sfrac{|\tau|}{T_L}}
\end{equation}

According to the Wiener-Khintchine theorem, the corresponding spectral power density $S(\nu)$ is then:
\begin{equation}
S(\nu) \propto \frac{1}{1+\left(\frac{2\nu}{\Delta\nu}\right)^2}
\end{equation}
where $\Delta\nu=1/(\pi T_L)$ is the full width at half maximum.

The coherence time is measured with two different techniques: Fabry-Perot interferometry and $\gtX_{\parallele}$ intensity correlation measurement at the output of the unbalanced Mach-Zehnder interferometer used for the Hong-Ou-Mandel (HOM) type experiment. The $\gtX_{\parallele}$ function is given for a lorentzian spectrum of coherence time $T_L$ and no amplitude fluctuations by \cite{cohChaoBeveratos}:
\begin{equation} \label{eq:laserG2}
\gtX_{\parallele}(\tau) = 1 -\frac12 \cdot  \big|g^{(1)}(\tau)\big|^2 = 1 - \frac12\cdot e^{-\sfrac{2 |\tau|}{T_L}}
\end{equation}

\begin{figure}[h!]
\captionsetup{width=0.95\linewidth}
\centering
\subfloat{\label{sfig:cohTimeComp}
\includegraphics{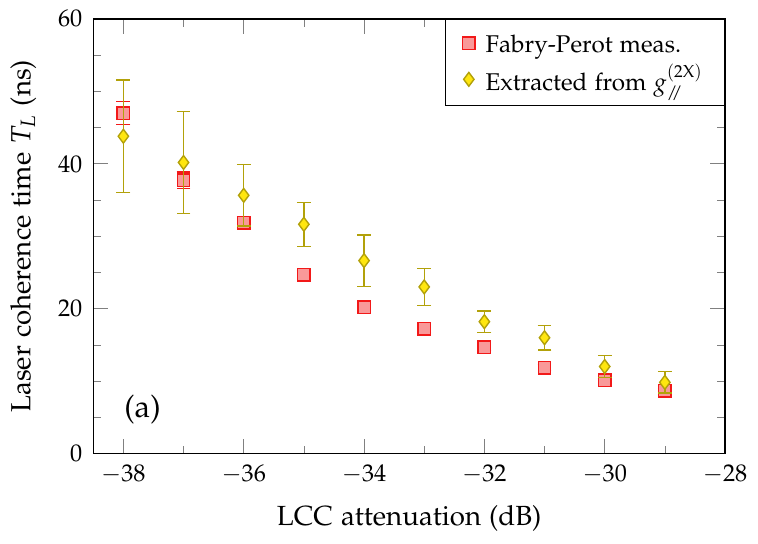}}
\sfigHspace
\subfloat{\label{sfig:homLaser}
\includegraphics{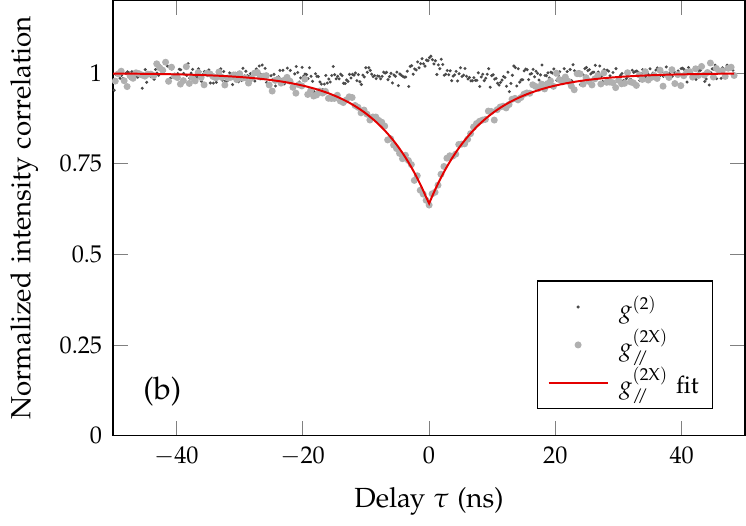}}
\caption{Effect of the LCC modulation attenuation on the coherence time of the excitation laser. \protect\subref*{sfig:cohTimeComp} Coherence time of the diode laser versus the LCC attenuation (i.e. the modulation power) extracted from the Fabry-Perot measurements (red squares, deconvoluted by the Fabry-Perot response) and the $\gtX_{\parallele}$ measurements (yellow diamonds). \protect\subref*{sfig:homLaser} Experimental data (light gray dots) and fit (solid red line) of the intensity correlation function $\gtX_{\parallele}$ when only the laser is sent to the HOM interferometer. The intensity correlation function $g^{(2)}$, measured in a HBT setup, is also shown (dark gray dots). The LCC attenuation is $-31\unit{dB}$ and the coherence time extracted from the fit is $T_L = 17.0\pm 0.5\unit{ns}$.}
\label{fig:VisibilityVSLinewidth}
\end{figure}

As shown in figure \ref{sfig:cohTimeComp}, these two techniques (red squares for Fabry-Perot and yellow diamonds for $\gtX_{\parallele}$) show only small differences. The coherence times in the main paper were measured using the $\gtX_{\parallele}$ method which is the most reliable one (the Fabry-Perot has a resolution of $10\unit{MHz}$ --- of the order of the narrowest laser linewidths --- and the measurement can then be altered for the highest values of the laser coherence time).

An example of a two-photon interference measurement performed on the laser only (i.e. $\gtX_{\parallele}$) is shown in figure \ref{sfig:homLaser} (light gray dots). The quality of the theoretical fit (red solid line) demonstrates the robustness of the $\gtX_{\parallele}$ method. However, a $g^{(2)}$ measurement performed in a Hanbury Brown and Twiss (HBT) setup exhibits a very small bunching (dark gray dots), of the order of $3\,\%$. This is linked to the LCC process, since the $g^{(2)}$ of a laser should read \cite{lasersSvelto}:
\begin{equation}
g^{(2)}(\tau) = 1 + \F[\text{RIN}]
\end{equation}
where $\F[\text{RIN}]$ is the Fourier transform of the laser Relative Intensity Noise. Here, because of the strong modulation on the diode laser current, the amplitude of the laser emission is also affected by the modulation, and the RIN goes from $10\unit{MHz}$ to $>250\unit{MHz}$, which would correspond to a bunching less than $8\unit{ns}$ wide. This bunching is in fact observed on the experimental results and is about $4\unit{ns}$ wide. Nevertheless, this effect is very small and the excitation laser can be considered as a quite stable field as far as the detection system is concerned. We also stress that, when measured on the QD, the single photon character of the QD emission attenuates considerably the bunching effect inherent to the excitation laser since it occurs on the same time scales. Thus, this phenomenon will affect very slightly the reliability of our model equations around zero delay. In this context, the bunching effect due to the laser modulation has been neglected in all the $g^{(2)}$ and $\gtX$ measurements presented in the paper.

\section{S2. One- and two-photon interference measurements}

The difference between one- and two-photon interference, also called second- and fourth-photon interference, is what is measured in the experiment, i.e. whether detecting one or two photons \cite{oneTwoMandel}. In the case of two-photon interference with a single source, the use of a strongly unbalanced interferometer prevents any constant phase correlation between its two outputs, or in other words it ensures the absence of one-photon interference. One could argue that the measurement of photon pairs is not altered by phenomena which modify the one-photon measurements. However, the individual rates fluctuations at the outputs of the beamsplitter will affect the visibility of the two-photon interference measurements, in addition to their influence on the coincidence rate.

\begin{figure}[h!]
\captionsetup{width=0.95\columnwidth}
\centering
\subfloat{\label{sfig:visVsLCC}
\includegraphics{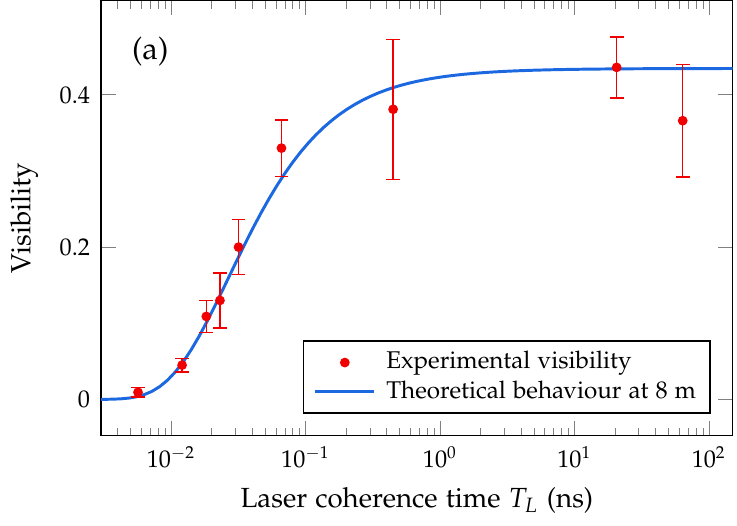}}
\hspace{0.5cm}%
\subfloat{\label{sfig:gtXQD}
\includegraphics{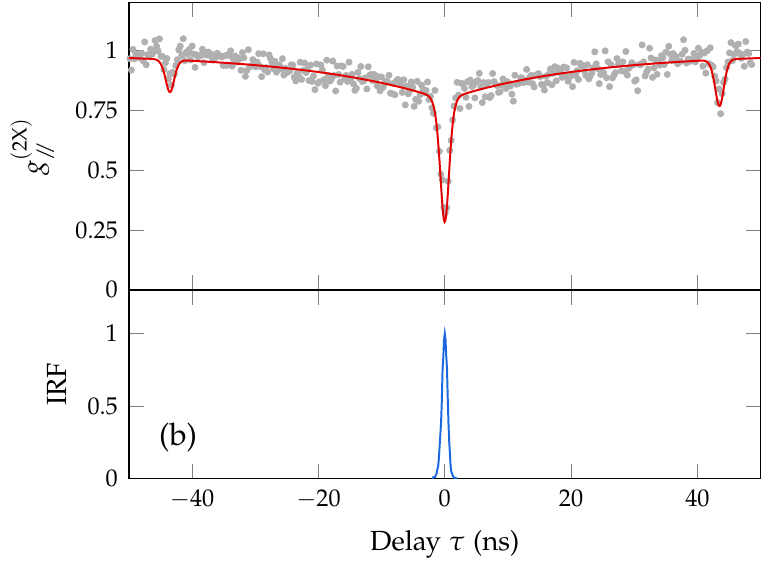}}
\caption{Condition on the HOM interferometer path difference for accurate two-photon interference measurements. \protect\subref*{sfig:visVsLCC} Visibility (red dots) of the one-photon interference fringes, measured in a Michelson interferometer with a delay of $27\unit{ns}$ (corresponding to $8\unit{m}$), as a function of the excitation laser linewidth. The blue solid line is the expected behavior for a lorentzian spectrum at such delay. \protect\subref*{sfig:gtXQD} Top: Intensity correlation function $\gtX_{\parallele}$ measured in the HOM type experiment for $T_L=43\unit{ns}$ (gray markers), fitted by Eq. 2 of the main paper (red solid curve). Bottom: Instrument Response Function of the HOM experiment (IRF, blue solid curve).}
\label{fig:interferenceAndDelay}
\end{figure}

Figure \ref{sfig:visVsLCC} shows the effect of the LCC modulation on the one-photon interference: the fringes visibility (red dots) decreases when the laser linewidth increases. This measurement was performed without fibers, with an unbalanced Michelson interferometer and a path difference of $27\unit{ns}$ (corresponding to 8 m). The result is consistent with the expected behavior of a lorentzian spectrum when varying the FWHM (blue solid line). Figure \ref{sfig:gtXQD} shows a $\gtX_{\parallele}$ measurement performed with the HOM interferometer on the QD emission (in the resonant Rayleigh Scattering regime) when the laser coherence time $T_L=43\unit{ns}$ is very close to the interferometer delay $\Delta \tau = 43.5\unit{ns}$. This results in a long decay which reaches the secondary peaks at $\pm \Delta \tau$. The experimental data are fitted by equation 2 of the main paper. We would like to stress that the model defined by equations 1-2 of the main paper cannot describe reliably the results of the HOM measurements as soon as the coherence time of the emitted photons gets longer than $\Delta\tau$. The detection system Instrument Response Function (IRF) of the HOM type experiment is also presented in figure \ref{sfig:gtXQD}.

\section{S3. Autocorrelation functions of a two-level system under resonant excitation}

For a two level system under resonant excitation, the first- and second-order correlation functions $g^{(1)}$ and $g^{(2)}$ are given by \cite{QO}:
\begin{equation}\label{g1}
g^{(1)}(\tau)=e^{-\sfrac{\tau}{T_L}}\left[
\frac{T_2}{2T_1}\cdot\frac{1}{1+\Omega^2T_1T_2}+
\frac{ e^{-\sfrac{\tau}{T_2}}}{2}+
\frac{e^{-\eta\tau}}{2}\left(\alpha\cos(\nu\tau)+\beta\sin(\nu\tau)\right) \right]
\end{equation}
and
\begin{equation}\label{g2}
g^{(2)}(\tau)=1-e^{-\eta\tau}\left(\cos(\nu\tau)+\frac{\eta}{\nu} \sin(\nu\tau)\right)
\end{equation}
with
\begin{equation}
\left\{
\begin{aligned}
\eta           &= \frac{1}{2}\left(\frac{1}{T_1}+\frac{1}{T_2}\right)\\
\nu            &= \sqrt {\Omega^2 -\frac{1}{4}\left(\frac{1}{T_1}-\frac{1}{T_2}\right)^2}\\
\alpha &= 1-\frac{T_2}{T_1\left(1+\Omega^2T_1T_2\right)}\\
\beta &= \frac{\Omega^2T_1\left(3T_2-T_1\right)-\frac{(T_1-T_2)^2}{T_1T_2}}{2\nu T_1\left(1+\Omega^2T_1T_2\right)}\\
\end{aligned} \right.
\end{equation}

\noindent $\tau$ is the delay between detections, $T_L$ the coherence time of the resonant excitation laser ($T_L>T_1,T_2$), and $\Omega$ the Rabi frequency defined as:
\begin{eqnarray}\label{Omega}
\Omega^2=\frac{P}{P_0}\frac{1}{T_1T_2}
\end{eqnarray}

\section{S4. Dynamical properties of the quantum dot emission}

\begin{figure}[h!]
\centering
\subfloat{\label{sfig:T1decay}
\includegraphics{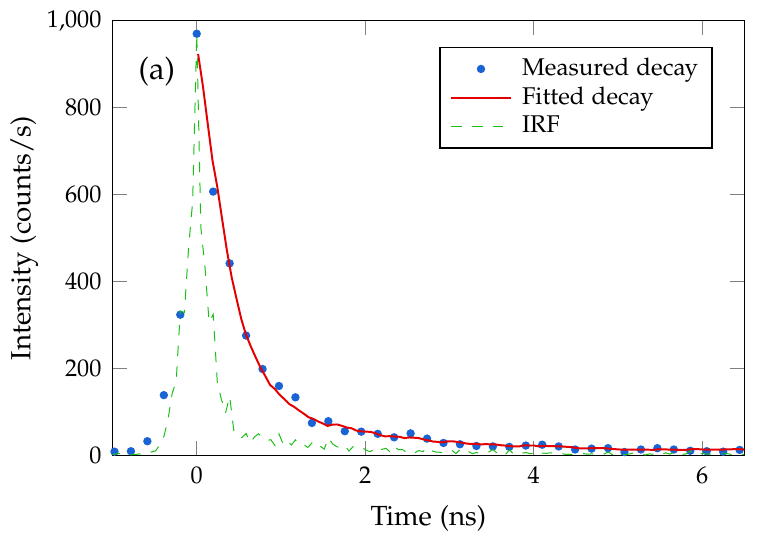}}\sfigHspace%
\subfloat{\label{sfig:T2mich}
\includegraphics{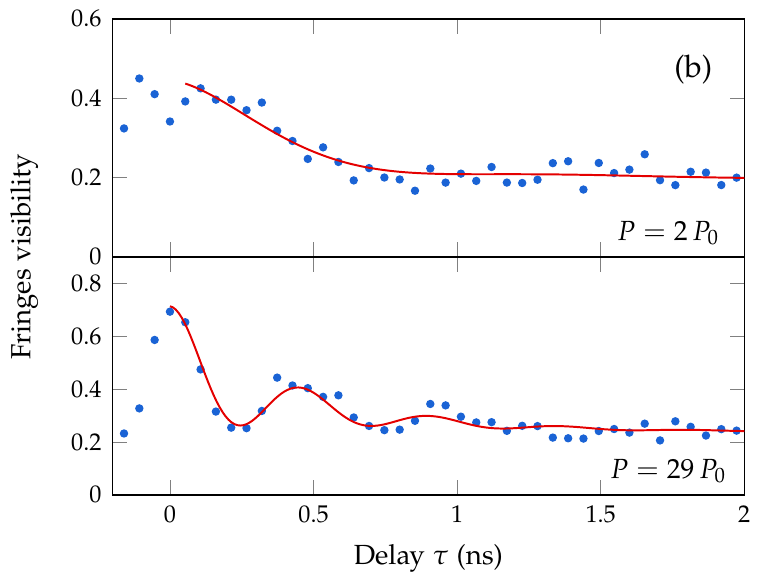}}
\caption{Dynamical properties of the quantum dot emission. \protect\subref*{sfig:T1decay} $T_1$ measurement: the experimental decay curve (blue dots) is presented with a fit (red solid line) of an exponential decay with a time constant $T_1 = 0.34 \pm 0.05 \unit{ns}$, convoluted by the Instrument Response Function (dashed green line). \protect\subref*{sfig:T2mich} $T_2$ measurement: fringes visibility as a function of the interferometer delay (blue dots), fitted by the $g^{(1)}$ function of a two-level system (Eq. 1) (red solid line). The excitation power is $P = 2\, P_0$ for the top panel and $P = 29\, P_0$ for the bottom panel. $T_1 = 0.3\unit{ns}$ for both. The extracted value of the coherence time is $T_2 = 0.5\pm 0.05\unit{ns}$.}
\label{fig:dynamics}
\end{figure}

Figure \ref{fig:dynamics} shows two experimental results used to assess the lifetime $T_1$ and the coherence relaxation time $T_2$ of the two-level system model that describes the QD. All the fits presented in the paper were done with the same experimental values of $T_1$ and $T_2$.

For $T_1$, a decay curve measurement was performed using a pulsed Ti:Sapphire laser tuned in resonance with the QD excitonic transition (see Fig.\,\ref{sfig:T1decay}), with a pulse width of the order of $10\unit{ps}$. The IRF (dashed green line) is presented and the IRF-convoluted exponential decay used for the fit (red solid line) implies a lifetime $T_1 = 0.34 \pm 0.05 \unit{ns}$.

For $T_2$, a $g^{(1)}$ measurement is performed in a balanced Mach-Zehnder interferometer by measuring the fringes contrast as a function of the interferometer delay (see Fig.\,\ref{sfig:T2mich}). The strong background of $20\,\%$ is due to residual scattered laser. The experimental results are fitted by the $g^{(1)}$ function of a two-level system, given by Eq.1, and the deduced coherence time is $T_2 = 0.5\pm 0.05\unit{ns}$. The two panels correspond to two different excitation powers. In particular, on the bottom panel, the Rabi oscillations that the two-level system undergoes when it is strongly coupled to the resonant laser above saturation are observable. The experimental value of $T_2$ reproduces accurately the data from the whole set of measurements performed for several excitation powers (not shown here). Moreover, the combination of the measured $T_1$ and $T_2$ describes reliably all the HBT and HOM measurements, raising a high level of confidence about these values.

\end{document}